\documentclass[conference]{IEEEtran}
\IEEEoverridecommandlockouts
\usepackage{hyperref}
\usepackage{amsmath,amssymb,amsfonts}
\usepackage{algorithmic}
\usepackage{graphicx}
\usepackage{textcomp}
\usepackage{xcolor}
\usepackage{todonotes}
\usepackage{tabularx, multirow}
\usepackage{fancyhdr,lipsum}
\usepackage{subfig}
\usepackage{float}
\fancypagestyle{firstpage}{
  \fancyhf{}
  \fancyfoot[l]{}
}

\def\BibTeX{{\rm B\kern-.05em{\sc i\kern-.025em b}\kern-.08em
    T\kern-.1667em\lower.7ex\hbox{E}\kern-.125emX}}
\usepackage[%
backend=biber,bibencoding=utf8, 
language=auto,
style=ieee,
sorting=none, 
maxbibnames=10, 
natbib=true 
]{biblatex}
\bibliography{main.bib}

\def\app#1#2{%
  \mathrel{%
    \setbox0=\hbox{$#1\sim$}%
    \setbox2=\hbox{%
      \rlap{\hbox{$#1\propto$}}%
      \lower1.1\ht0\box0%
    }%
    \raise0.25\ht2\box2%
  }%
}
\begin{document}
\title{ Offshore Wind Integration in the North Sea: \\ The Benefits of an Offshore Grid and Floating Wind

}
\author{
	\IEEEauthorblockN{1\textsuperscript{st} Philipp Glaum}
	\IEEEauthorblockA{\textit{Institute of Energy Technology} \\
		\textit{Technische Universität Berlin}\\
		Berlin, Germany \\
		p.glaum@tu-berlin.de}
	\and
	\IEEEauthorblockN{2\textsuperscript{nd} Fabian Neumann}
	\IEEEauthorblockA{\textit{Institute of Energy Technology} \\
		\textit{Technische Universität Berlin}\\
		Berlin, Germany \\
		f.neumann@tu-berlin.de}
	\and
	\IEEEauthorblockN{3\textsuperscript{rd} Tom Brown}
	\IEEEauthorblockA{\textit{Institute of Energy Technology} \\
		\textit{Technische Universität Berlin}\\
		Berlin, Germany \\
		t.brown@tu-berlin.de}
}
\maketitle
\thispagestyle{firstpage}

\begin{abstract}

Wind energy has become increasingly important for meeting Europe's energy needs.
While onshore wind expansion faces public acceptance problems, for offshore wind the European Commission has introduced ambitious goals to increase capacity from 15~GW to 300~GW in 2050.  Incorporating more offshore wind electricity into the power grid may offer a more widely accepted way to satisfy Europe's energy demand.  In particular, the North Sea region has large potential for offshore wind generation.  However, to fully exploit the wind potential in the North Sea, the grid integration of offshore wind and floating wind turbines are vital, especially when onshore wind capacity and onshore grid expansion are constrained.  For the grid integration, a meshed offshore grid can offer a viable alternative to the standard direct connection of offshore wind parks to the nearest point on land combined with point-to-point HVDC connections.  In this paper we investigate the benefits of having a meshed offshore grid in the North Sea and considering floating wind besides fixed-bottom wind installations.  In our analysis, we look at eight different scenarios, where onshore wind potentials and onshore line expansion are limited, to explore the effects of low public acceptance.  Our results demonstrate that the presence of an offshore grid can reduce total system costs by up to 2.6~bn€/a.  In the scenarios with an offshore meshed grid $\sim$8\% more offshore wind capacities are built compared to the scenarios without a meshed grid.  Furthermore, the analysis shows that if onshore wind potentials are restricted, floating wind turbines play a key role and compensate for lacking onshore wind capacities.
\end{abstract}


\section{Introduction}\label{sec:introduction}
The European Commission has outlined a plan to significantly increase the capacity of offshore wind power generation in Europe from 15~GW in 2021 to approximately 300~GW in 2050 \cite{europeancommissionEUOffshoreRenewable2020}.
To supply Europe's energy needs, the demand for offshore wind turbines is further exacerbated by a decreasing viability of onshore wind power due to and growing opposition from local communities and less favourable wind conditions \cite{hevia-kochComparingOffshoreOnshore2019}.
Especially the North Sea Region (NSR) with its rich offshore wind generation potential is expected to play a pivotal role in achieving these expansion targets, and is expected to lead Europe's continental energy transition \cite{martinez-gordonModellingHighlyDecarbonised2022}.
It is estimated that the NSR has a sizeable potential of more than 635~GW for offshore wind generation capacities \cite{europeancommission.directorategeneralforenergy.HybridProjectsHow2019}.
Furthermore, NSR is surrounded by multiple countries with a joint population of 200~million and energy-intensive industries accounting for roughly 60\% of Europe's Gross Domestic Product (GDP)  \cite{martinez-gordonModellingHighlyDecarbonised2022}.
As part of the North Sea Energy Cooperation, the European Union (EU) has set ambitious targets to build at least 260~GW of offshore wind energy by 2050 and 76~GW by 2030 in the NSR \cite{northseasenergycooperationJointStatementNorth2022}.
While currently fixed-bottom turbines are the norm in the offshore wind industry, considering floating wind turbines could unlock further potentials in the NSR aiding the energy transition to net-zero emissions.
Floating wind turbines represent the new frontier in the offshore wind industry \cite{maienzaLifeCycleCost2020}.
While floating wind installations in Europe are still rare, the market for this technology is rapidly emerging, with many projects planned to be operational by 2030 \cite{diazMarketNeedsOpportunities2022}.
Floating wind turbine has the potential to unlock resources at previously inaccessible water depths, further expanding the potential for offshore wind power generation \cite{cruzFloatingOffshoreWind2016}.
Besides accessing the offshore wind potential, an efficient integration the offshore wind energy is essential.
Consequently, the EU recognizes the development of offshore wind integration and offshore grid infrastructure in the North Sea a long-standing priority \cite{entso-eOffshoreGridDevelopment2011}.

Current national approaches to integrate offshore wind farms involve connecting each farm with point-to-point connections designed to transport the maximum output of the wind farm directly to designated onshore locations.
However, this approach results in partially unused connections when the farm operates below its maximum line capacity.
The connection of multiple offshore wind farms in a meshed offshore power grid could offer an alternative to direct point-to-point connections.
An offshore grid may offer several additional benefits like (1) network resilience for wind integration, (2) increased capacity for cross-border trade, (3) fewer, larger assets which can reduce overall capital cost \cite{entso-eENTSOEPositionOffshore2020}.
Moreover, the inclusion of offshore energy hub facilities can increase the offshore wind capacity and, when combined with hydrogen production, can reduce the curtailment of offshore wind \cite{durakovicPoweringEuropeNorth2022}.

The North Sea Wind Power Hub, a consortium of three transmission system operators (TSOs) from North Sea countries, seeks to play a major role in developing integrated systems in the North Sea with offshore hubs \cite{northseawindpowerhubconsortiumHubsSpokesViable2022}.
Their project has been granted the status of a Project of Common Interest by the European Union, and it has confirmed the technical feasibility of an offshore power hub and grid in the North Sea.
However, the deployment of offshore projects in the North Sea still faces several barriers, including planning across assets and countries, and uncertainty about responsibility regarding maritime space \cite{europeancommission.directorategeneralforenergy.HybridProjectsHow2019}.

Several studies have investigated the integration of offshore wind in the North Sea region (NSR) with an offshore grid, using energy optimization models or simulations to analyze the benefits of such a grid \cite{gea-bermudezOptimalGenerationTransmission2020,martinez-gordonBenefitsIntegratedPower2022,durakovicPoweringEuropeNorth2022,houghtonOffshoreTransmissionWind2016}.
All of these studies demonstrate an increased offshore wind capacities built when an offshore grid can be leveraged.
For instance, the analysis by Gea-Bermúdez et al.\cite{gea-bermudezOptimalGenerationTransmission2020} obtained total system cost savings of approximately 1~bn€ per year for the model incorporating an offshore grid.
Moreover, in their analysis, the offshore grid leads to the integration of more offshore wind capacities, with less onshore wind and solar PV capacities needed.
In Durakovic et al.\cite{durakovicPoweringEuropeNorth2022}, the authors observe even stronger increases in installed offshore wind capacity with the aid of an offshore grid, particularly with on-site hydrogen production.

However, none of these studies have included an endogenously optimization of offshore grid topology in high spatial detail while considering both fixed-bottom and floating wind turbines.
Our paper aims to fill this gap by presenting a pan-European energy system analysis of the cost benefits of an offshore grid in the NSR in high spatial and temporal resolution with floating wind.

\section{Methodology} \label{sec:methodology}
In this section, we present the methodology on which model we are utilizing and how we build the model to investigate the benefits of floating wind and an offshore grid in the North Sea.

\subsection{Energy System} \label{sec:energy_system}

Our analysis is based on the open European energy system model PyPSA-Eur-Sec \cite{neumannPyPSAPypsaeursecPyPSAEurSec2023}. This optimization model is designed to minimize both investment and operation costs while taking into account techno-economic constraints including linearized power flow equations, wind and solar availability and potentials, and emission limits. The model encompasses all energy sectors, namely industry, transport, energy, agriculture, and buildings. In the Appendix \ref{sec:offwind_technology}, we provide the modelling parameters for the offshore wind technologies.

\subsection{Enhanced modelling of offshore wind potentials}\label{sec:offshore_modelling}

\begin{figure}[htb]
	\centering
	\includegraphics[width=0.4\textwidth]{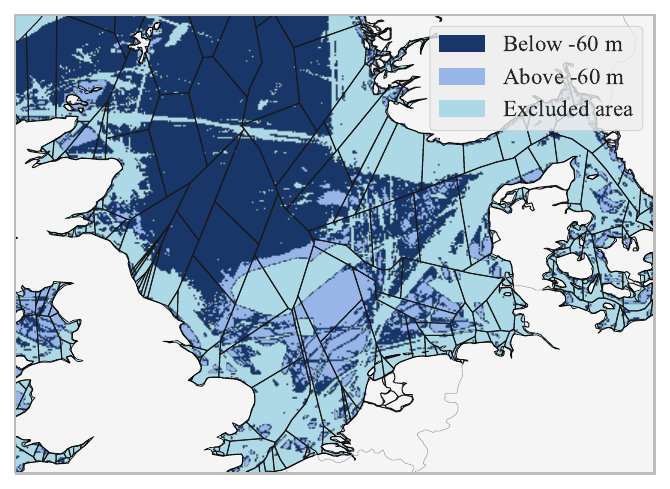}
	\caption{Eligible areas (blue) for offshore wind capacities in the North Sea region with exclusion zones for nature reserves and busy shipping lanes (white). The black polygons denote the offshore regions as passed to the capacity expansion model.}
	\label{fig:exclusion area}
\end{figure}

To improve the accuracy and reliability of how offshore wind potentials are modelled, we have extended the PyPSA-Eur-Sec framework in several ways.
The first extension (1) implements the inclusion of floating wind technology besides fixed-bottom technologies.
This extension addresses the growing importance of floating wind technology in the offshore wind energy sector, especially when the feasible potential for onshore wind is constrained.
The second extension (2) increases the resolution of offshore regions based on the countries' exclusive economic zones (EEZ).
An offshore region denotes a particular area for which hourly wind generation profiles and total capacity potentials are calculated. For each region, the model can choose how much wind capacity is built up until the potential limits.
Previous model versions represented offshore regions by elongated shapes based on Voronoi cells to the closest onshore region.
In this paper, these regions have been divided into smaller elements, allowing for a higher resolution of offshore potential and better differentiation between near-shore and far-shore sites.
The subdivision of offshore regions works in a way that the economic exclusive zones of every region retain a direct allocation of the wind resources to the corresponding country.
The third extension (3) improves the calculation of the areas eligible for offshore wind park development.
While previous model versions already excluded natural protection areas and limits to the water depths for fixed-bottom developments, we now also exclude busy shipping lanes.
The outcomes of extensions (2) and (3) are illustrated in Fig. \ref{fig:exclusion area}.
The fourth expansion (4) improves the cost modelling of the fixed-bottom technologies.
Previous model versions applied constant investment costs for fixed-bottom installations regardless of water depth and turbine type.
For our improvements, we adopt the cost model developed by the Danish Energy Agency \cite{danishenergyagencyTechnologyData2022}, which considers the water depth, rotor diameter, hub height, specific power output and wind farm size to determine site-specific investment costs for fixed-bottom wind turbine installations.

\subsection{Enhanced modelling of offshore grid topology}\label{sec:offshore_grid}

\begin{figure}[h]
	\centering
	\includegraphics[width=0.5\textwidth]{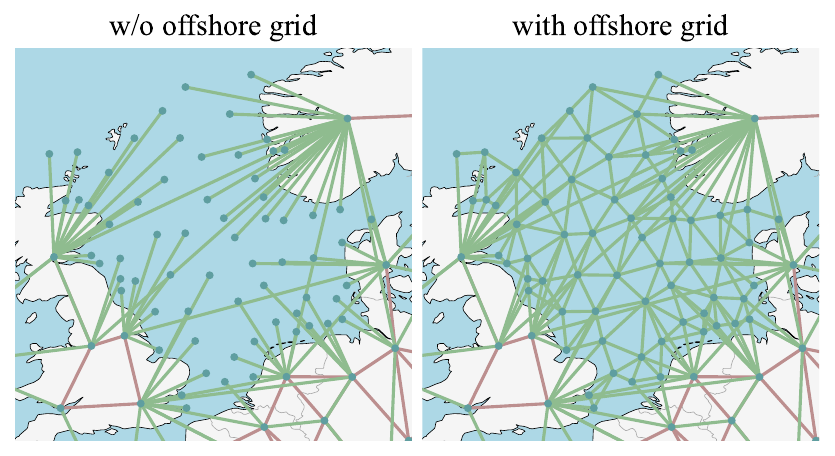}
	\caption{Extract of the possible offshore grid topology in the NSR. Lines only reflect the decision space of the optimization model.}
	\label{fig:offgrid_input}
\end{figure}

This section explains how we model options for the development of new offshore grid topology.
The capacity of each connection option is a decision variable, such that the model can
endogenously decide for the most cost-effective options to integrate offshore wind into the European energy system.
The offshore regions introduced in Section \ref{sec:offshore_modelling} determine the offshore grid topology in the model.
All regions can connect directly to the closest onshore bus through point-to-point connections in all scenarios.
Scenarios with the option for developing a meshed offshore grid additionally allow for new transmission lines between adjacent regions.
Thereby, these scenarios gain an additional degree of freedom to interconnect the offshore regions, allowing greater flexibility in optimizing the offshore grid and consideration of specific characteristics.
Fig. \ref{fig:offgrid_input} shows the grid expansion options in the NSR for both scenarios.
The initial capacity of offshore transmission lines is zero. For the development of a meshed offshore grid we assume controllable HVDC lines; power flows are represented as transport model, whereas
linearized power flow equations are applied to existing onshore transmission capacities.
Even though our paper focuses on offshore grid integration in the North Sea, our method is also applicable to other regions.

\section{Case Studies}\label{sec:study-cases}

Applying the methodology described in Section \ref{sec:offshore_modelling},
the model comprises 64 onshore regions and 66 offshore regions and has a 3-hourly temporal resolution across a full year.
To investigate the potential benefits of the offshore grid and floating wind, we vary three parameters to create eight different scenarios.
The first parameter, \textbf{offshore grid}, determines whether a meshed offshore grid topology may be developed.
The second parameter, \textbf{transmission limit}, specifies the total expansion limit for transmission line relative to the current transmission network measured in MWkm.
For instance, for a transmission limit of 100\% restricts the power transmission capacities to today's levels.
Because this sensitivity addresses the potential of limited public acceptance for overhead transmission lines, we do not constraint the expansion of underground or submarine cables.
Therefore, the model can always expand the offshore grid transmission lines.
The third parameter, \textbf{onshore wind potential}, represents the available onshore wind potential in the model relative to the total socio-technical onshore wind potential.
For example, 25\% onshore wind potential means that only a fourth of the total onshore wind potential may be built.
This sensitivity accounts for the limited support the development of onshore wind turbines experiences in many countries in Europe such as France and the UK.
The selected values for the different parameters in the scenarios are given in Table \ref{tab:scenario_parameters}.

\begin{table}
	\caption{Scenario definition.}
	\begin{tabularx}{0.49\textwidth}{X|X|c|c}
		Parameters & Offshore grid & Transmission limit & Onshore wind potential \\
		\hline
		Values     & with grid     & 100\% transmission & 25\% onshore wind       \\
		           & w/o grid      & 150\% transmission & 100\% onshore wind      \\
	\end{tabularx}
	\label{tab:scenario_parameters}
\end{table}

As we analyze all possible combinations of the parameters from Table \ref{tab:scenario_parameters}, we obtain a total of eight different scenarios.
All scenarios represent overnight scenarios in the year 2050 considering a fully decarbonized energy system taking cost assumptions for 2030.
With the different scenarios, we aim to analyze how the offshore grid and floating wind turbines can help to relieve limited transmission capacities and limited onshore wind.
However, the 100\% onshore wind potential included in our model only reflects the technical potential. National governments typically have onshore wind targets closer to 25\% than 100\% of that value.
For example, Germany currently has the most onshore wind capacity installed in Europe and plans to have 115~GW installed by 2030 \cite{federalministryofeconomicaffairsandclimateactionUeberblickspapierOsterpaket2022}.
The technical potential in our model is 490~GW, which is highly unlikely to be reached by 2050 due to social acceptance issues \cite{hevia-kochComparingOffshoreOnshore2019}.
\section{Results}\label{sec:results}

\begin{figure}[h]
	\centering
	\includegraphics[width=0.5\textwidth]{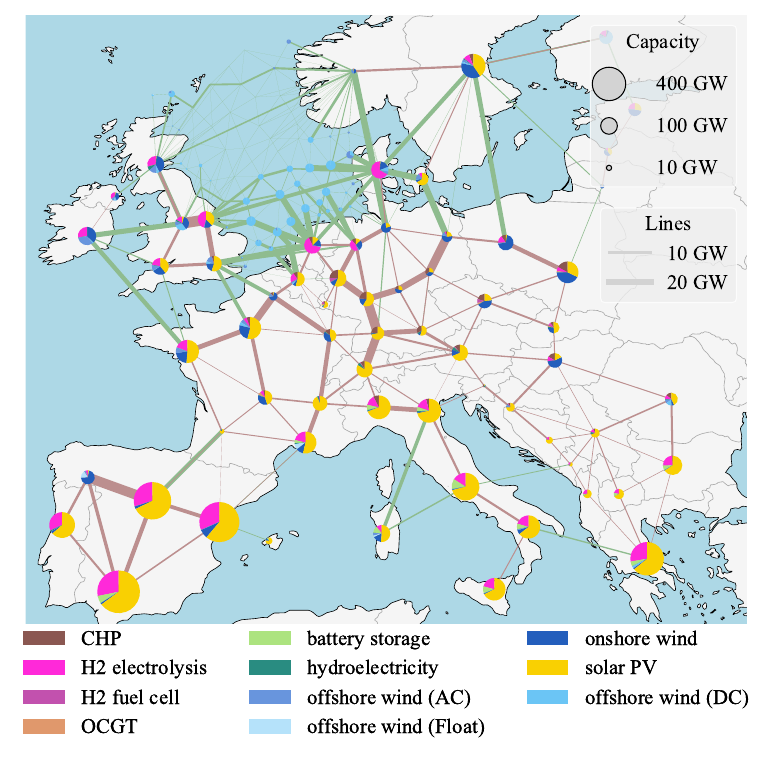}
	\caption{Capacity map for the scenario with grid, 100\% transmission and 25\% onshore wind.}
	\label{fig:grid_onwind0.25_lv1.0}
\end{figure}

We first present our results of the most restricted scenarios with 25\% onshore wind potential and no transmission expansion (100\% transmission limit).
For this case, Fig.~\ref{fig:grid_onwind0.25_lv1.0} shows a map of the generation, storage, conversion and transmission capacities built.

The map reveals that especially in the South of the NSR the model builds offshore transmission to improve the connections to the United Kingdom (UK), Denmark, the Netherlands, and Germany.
Despite the presence of direct connections between Denmark and the UK, the model builds additional transmission lines to collect the power output of multiple offshore wind farms and reinforce cross-border transmission between North Sea countries.
Additionally, a route between the far North of the UK and Norway is built.
Moreover, we observe that floating wind capacities are not built in the NSR.
However, they appear in different locations, such as north of Spain, south of France and around the UK.
This is because most available sites for offshore wind in the NSR are in water depths below 60m, where it is still feasible and less costly to deploy fixed-bottom turbines.

\begin{figure}[h]
	\centering
	\includegraphics[width=0.5\textwidth]{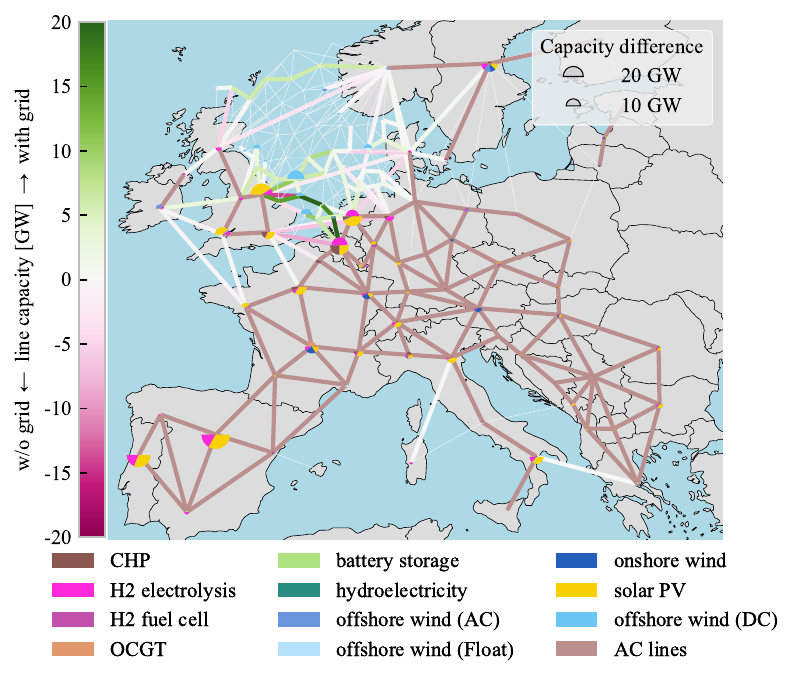}
	\caption{Comparison of capacities built between scenarios with and w/o grid, and 100\% transmission and 25\% onshore wind. Positive colored labels and upper semicircles represent a higher built out in the grid scenario.}
	\label{fig:comparison_onwind0.25_lv1.0}
\end{figure}

In Fig.~\ref{fig:comparison_onwind0.25_lv1.0}, we compare the capacity difference of the offshore grid scenario with 25\% onshore potential and 100\% line limit with the scenario without grid.
The upper semicircle indicates a higher build-out in the offshore grid case compared to the without grid case, and the lower semicircle vice versa.
The color bar of the offshore lines shows which lines are expanded more with a meshed offshore grid (red) and which are expanded less (green).

Note here that only the offshore grid case can build out interconnections of the offshore buses, i.e. the interconnections can only be colored in red, and that the onshore transmission lines are not expanded.
In terms of electricity generation capacities, allowing a meshed offshore grid entails more offshore wind capacities in the NSR and, consequently, substitutes some onshore wind and solar PV capacities.
Furthermore, more electrolysers are built in close proximity to the NSR in the case with a meshed offshore grid.
On the other hand, in the case without an offshore grid, more electrolysers are built in sunny regions like Spain and Portugal.

\begin{figure}[h]
	\centering
	\includegraphics[width=0.5\textwidth]{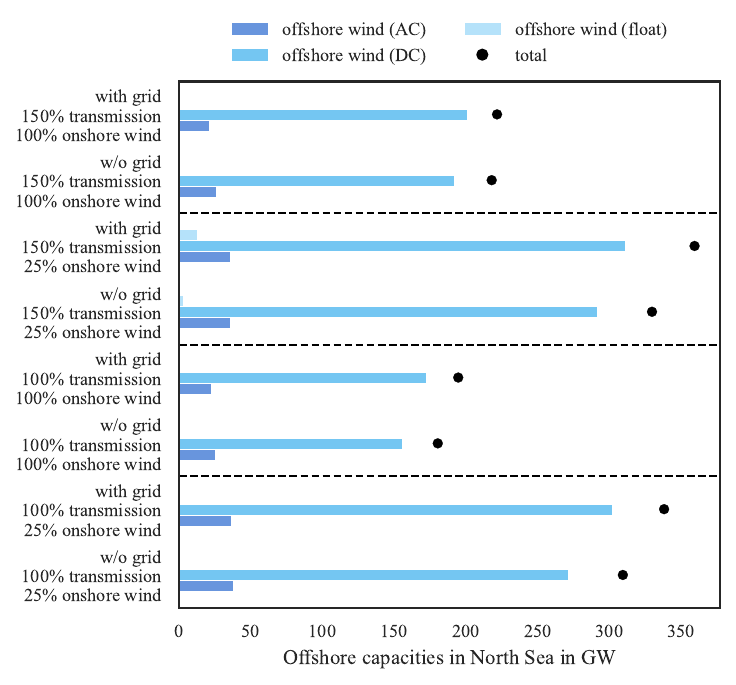}
	\caption{Comparison of offshore wind capacities in the North Sea region for the different scenarios.}
	\label{fig:offind_capacities}
\end{figure}

In Fig.~\ref{fig:offind_capacities}, we take a closer look at the offshore wind integration in the NSR for the given scenarios.
We observe a general increase in offshore wind capacities due to the offshore grid in all corresponding scenarios.
However, the build-out rate of offshore wind highly depends on the restrictions placed on the onshore wind potential.
Especially, when the expansion of onshore wind turbines is limited, the models build roughly twice as much offshore wind capacities.
Floating wind is only built in the NSR when the onshore wind is limited to 25\% and the offshore grid exists.
For the grid scenario with 25\% onshore wind and 150\% transmission, more offshore wind capacities are built than in the scenario without onshore grid reinforcements.
A possible explanation is that the presence of more onshore transmission capacities enable the transport of more electricity from offshore sites to inland demand centers.
The increase of offshore wind capacities in the NSR for the grid scenarios compared to the no grid scenarios ranges from 7.9\% to 9.3\%.

\begin{figure}[h]
	\centering
	\includegraphics[width=0.5\textwidth]{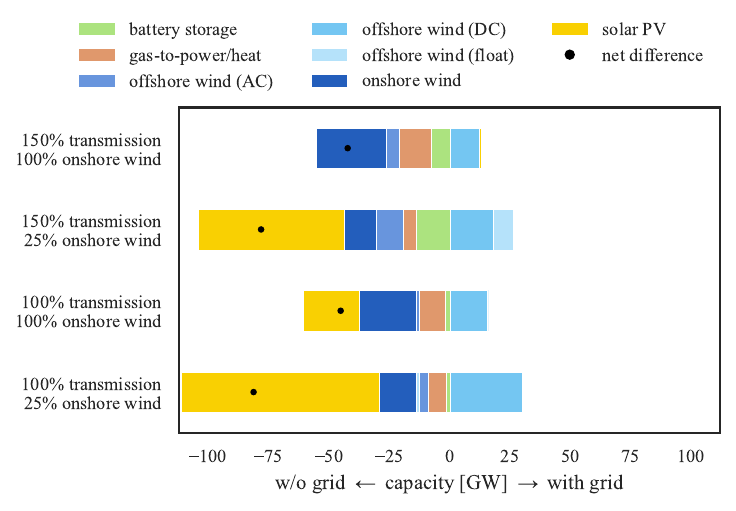}
	\caption{Generation capacity comparison of the scenarios considering offshore grid to without grid.}
	\label{fig:capacity_comparison}
\end{figure}

In Fig. \ref{fig:capacity_comparison}, we compare the mix of generation capacities between the scenarios with a meshed offshore grid and with exclusive point-to-point connections.
Positive numbers indicate a higher expansion in the grid case and vice versa.
It is prominent that all scenarios with offshore grid exhibit lower generation capacities overall.
This can be attributed largely to higher capacity factors of offshore wind compared to onshore wind and solar, and reduced balancing need because of more stable production patterns.
In general, in the grid scenarios offshore wind capacities replace onshore wind, solar PV, gas-to-power/heat and battery discharge capacities.
Particularly, in the scenarios with 25\% onshore wind, the offshore grid demonstrates an efficient integration of offshore capacities.
Moreover, we observe that the model prefers far-shore, DC-connected offshore wind in scenarios with an offshore grid, whereas near-shore, AC-connected offshore wind is more prevalent in scenarios without it.
This finding suggests that the offshore grid facilitates the cost-effective integration of remote offshore sites with abundant wind resources.

\begin{figure}[h]
	\centering
	\includegraphics[width=0.5\textwidth]{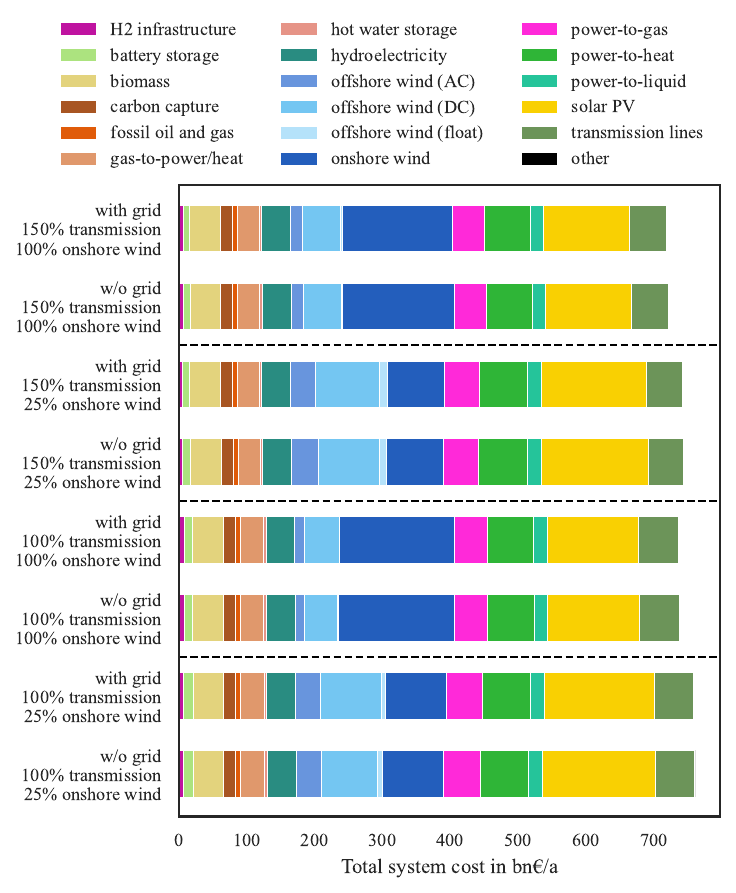}
	\caption{Comparison of total energy system costs by component between the different scenarios.}
	\label{fig:total_system_cost}
\end{figure}

Fig. \ref{fig:total_system_cost} presents the total system costs of the different scenarios.
The cost savings for the scenarios with an offshore grid compared to those without grid range from 1.7 to 2.6~bn€/a, where those with 150\% transmission and 25\% onshore wind yields the highest cost savings.
However, the cost benefit has a limited role with 0.4\% when related to the cumulative total system cost.
It is worth noting that the main impact on the deployment of floating wind turbines is not the existence of an offshore grid but the lack of onshore wind potential.
This is due to the fact that with missing onshore wind potentials the need for further energy resources offshore arises and the resources in shallow waters below a depth of 60m are already exhausted.
Therefore, only floating wind can compensate for the lack of onshore wind potentials. %

Additional plots for the cumulative installed capacities and the curtailed energy are given in Appendix \ref{sec:appendix_plots} in Fig. \ref{fig:appendix_capacity_comparison} and \ref{fig:appendix_curtailment_comparison} .

\section{Limitations}\label{sec:Limitations}
Limitations of this study include that the offshore grid topology used in the analysis is highly meshed and may not reflect the current plans of transmission network operators.
Other studies like \cite{gea-bermudezOptimalGenerationTransmission2020,durakovicPoweringEuropeNorth2022, martinez-gordonBenefitsIntegratedPower2022}, only consider a few offshore nodes or power hubs to model the offshore grid.
Furthermore, this study models the offshore grid only with HVDC lines and does not consider the use of HVAC lines.
However, offshore grids typically utilize a cost-optimal mix of AC and DC connections depending on the line lengths.
Moreover, the optimization model used in this study is a linear programming problem and does not allow for discrete investment decisions, which may not accurately capture the investment behavior of real-world decision makers. Owing to the linearization we also do not consider transmission losses and simplify nonlinear power flows.

\section{Conclusion}\label{sec:conclusion}

In summary, this paper investigates the advantages of an offshore grid in the North Sea region and the potential of floating wind turbines through eight distinct scenarios.
Our results demonstrate that the existence of an offshore grid promotes the cost-effective integration of offshore wind and decreases the necessity for onshore wind, solar PV, and gas-to-power/heat.
Furthermore, more electrolyser units are built in proximity to the NSR and replace units in sunny regions like Spain.
This allows for the production of hydrogen closer to demand centers in the NSR, which consists of highly industrialized countries.
Our model demonstrates total system cost savings of up to 2.6bn€/a with the presence of an offshore grid.
Notably, the benefits of the offshore grid are most significant when onshore transmission capacities are enhanced.
This suggests that the benefits of an offshore grid are impaired when onshore wind capacities cannot be expanded to transmit electricity further to load centers onshore.
Our study further indicates a critical role of floating wind turbines, particularly when onshore wind capacities are limited.
In this case, floating wind supplements fixed-bottom wind turbines and is deployed with capacities up to 45~GW.

\newpage

\printbibliography
\newpage
\appendix
\subsection{Offwind technology data}\label{sec:offwind_technology}
As offshore wind turbine type we use the National Renewable Energy Laboratory (NREL) 12 MW offshore reference turbine from Table \ref{tab:12MW_turbine} \cite{nrelNRELTurbineModels}.
\begin{table}[H]
	\centering
	\caption{Key figures for the NREL 12 MW offshore reference turbine.}
	\begin{tabular}{l|l|l}
		Item               & Value & Units \\
		\hline
		Rated power        & 12    & MW    \\
		Cut-in wind speed  & 4     & m/s   \\
		Cut-out wind speed & 25    & m/s   \\
		Rotor diameter     & 214   & m     \\
		Hub height         & 136   & m     \\
	\end{tabular}
	\label{tab:12MW_turbine}
\end{table}

Table \ref{tab:offshore_parameters} shows the modelling parameters for the offshore wind technologies incorporated in the model.
\begin{table}[H]
	\centering
	\caption{Parameters for offshore wind technologies.}
	\begin{tabularx}{0.49\textwidth}{l|X|X|X}
		Technology       & Distance to shore & Water depth & Capacity density          \\
		                 & [km]              & [m]         & [$\text{MW}/\text{km}^2$] \\
		\hline
		Offwind AC       & 10-30             & $<$60       & 5                         \\
		Offwind DC       & $>$30             & $<$60       & 5                         \\
		Offwind Floating & $>$10             & $>$60       & 5                         \\
	\end{tabularx}
	\label{tab:offshore_parameters}
\end{table}

\subsection{Further Scenario Plots}
\label{sec:appendix_plots}
\begin{figure}[h]
	\includegraphics[width=0.5\textwidth]{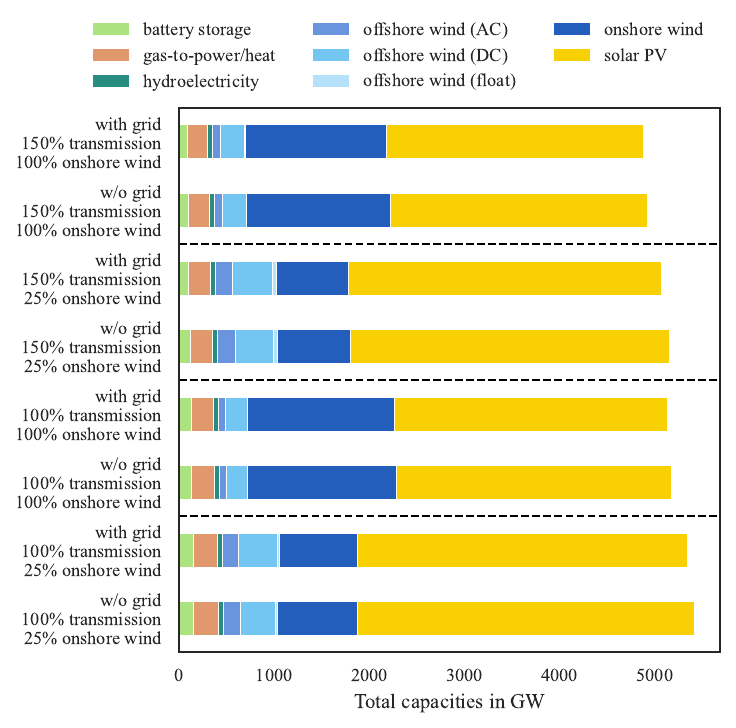}
	\caption{This figure shows the total generation capacity in GW for all 8 scenarios. Prominently, the more restrictions we experience regarding onshore transmission capacities and onshore wind potentials, the more generation capacities the model deploys.}
	\label{fig:appendix_capacity_comparison}
\end{figure}

\begin{figure}[H]
	\includegraphics[width=0.5\textwidth]{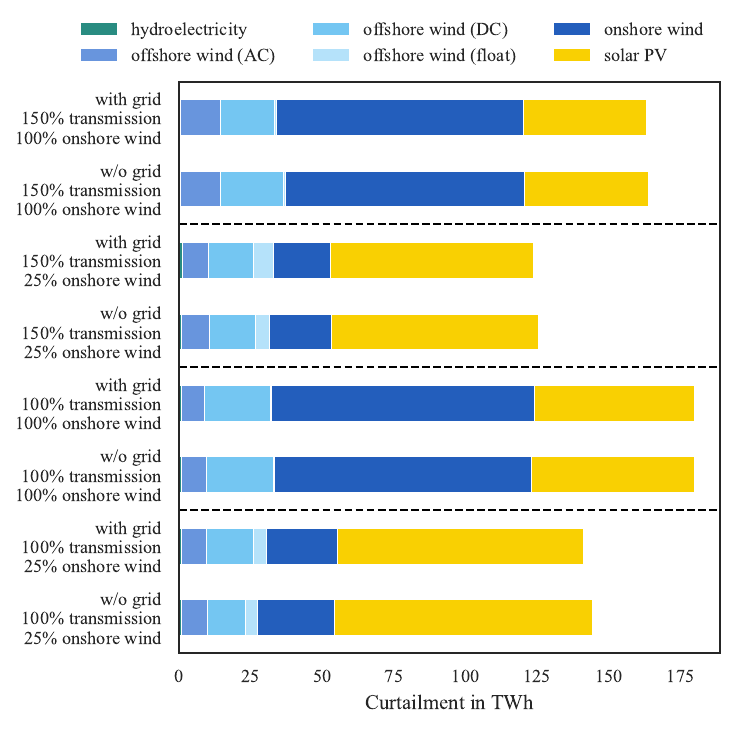}
	\caption{This figure shows the curtailed electricity in TWh for all 8 scenarios. We observe that curtailment reduces with increasing transmission capacities and with the absence of an offshore grid. The reason for this is that electricity can be better transmitted to demanders. However, in our scenarios excess electricity accounts for the major reason of curtailment.}
	\label{fig:appendix_curtailment_comparison}
\end{figure}
\end{document}